\documentclass[aps,prb,twocolumn,showpacs,10pt,superscriptaddress]{revtex4-1}

\usepackage{bm}
\usepackage{graphicx}
\usepackage{amssymb}
\usepackage{amsmath}

\begin{document}

\title{Bound states in the continuum driven by AC fields}

\author{C.\ Gonz\'{a}lez-Santander}
\affiliation{GISC, Departamento de F\'{\i}sica de Materiales, Universidad
Complutense, E-28040 Madrid, Spain}

\author{P.\ A.\ Orellana}
\affiliation{Departamento de F\'{\i}sica, Universidad Cat\'{o}lica del Norte,
Casilla 1280, Antofagasta, Chile}

\author{F. Dom\'{i}nguez-Adame}
\affiliation{GISC, Departamento de F\'{\i}sica de Materiales, Universidad
Complutense, E-28040 Madrid, Spain}

\begin{abstract}

We report the formation of bound states in the continuum driven by AC fields.
This system consists of a quantum ring connected to two leads. An AC side-gate
voltage controls the interference pattern of the electrons passing through the
system. We model the system by two sites in parallel connected to two
semi-infinite lattices. The energy of these sites change harmonically with time.
We obtain the transmission probability and the local density of states at the
ring sites as a function of the parameters that define the system. The
transmission probability displays a Fano profile when the energy of the incoming
electron matches the driving frequency. Correspondingly, the local density of
states presents a narrow peak that approaches a $\delta$ function in the weak
coupling limit. We attribute these features to the presence of bound states in
the continuum.

\end{abstract}


\maketitle

\section{Introduction}

At the dawn of quantum mechanics, von Neumann and Wigner constructed a spatially
oscillating attractive potential that supported a bound state above the
potential barrier.\cite{Neumann29} This truly localized (square integrable)
solution of the time-independent Schr\"{o}dinger equation is referred to as a
bound state in the continuum~(BIC). Much later, Stillinger and Herrick
reexamined and extended von Neumann and Wigner ideas.\cite{Stillinger75} They
analyzed a double excited atom model, where BICs were formed and live forever
despite the interaction between electrons. They arrived at the conclusion that
BICs may be a physically realizable phenomenon in real atomic and molecular
systems. In this context, Friedrich and Wintgen discussed in a system of coupled
Coulombic channels and, in particular, a hydrogen atom in a uniform magnetic
field.\cite{Friedrich85} These authors interpreted the formation of BICs as the
result of the interference between resonances of different channels. The
discovery of a BIC induced by the interaction between two particles in close
proximity to an impurity has been recently reported by Zhang~\emph{et
al.\/},\cite{Zhang12} where the state can be tuned in and out of the continuum
continuously.

The advent of nanofabrication techniques has made it possible to devise and
fabricate quantum devices whose electronic properties are similar to those of
atoms and molecules. When the size of the device is comparable to the de Broglie
wavelength, one or more degrees of freedom are quantized and the  electron wave
function is spatially confined. The similarity to atomic systems paved the way
to experimentally validate the existence of BICs in artificial nanostructures.
Capasso~\emph{et al.\/} measured the absorption spectrum at low temperature of a
GaInAs quantum well with Bragg reflector barriers produced by a AlInAs/GaInAs 
superlattice.\cite{Capasso92} A well defined line at $360\,$meV in the spectrum
was attributed to electron excitations from the ground state of the quantum well
to a localized level well above the AlInAs band edge. Nevertheless, this state
cannot be regarded as a true BIC but a bound state above the barrier since it is
a defect mode residing in the minigap of the superlattice, as pointed out by
Plotnik \emph{et al.}\cite{Plotnik11} Thus, the experimental validation of the
existence of truly BICs in quantum systems still remains a challenging task.
Furthermore, it is worth to mention in this context that the analogy between
photonic systems in the paraxial regime and quantum systems has facilitated the
study~\cite{Longhi07,Moiseyev09,Bulgakov10} and subsequent experimental
observation of BICs.\cite{Plotnik11}

Electronic transport in mesoscopic and nanoscopic systems can be also influenced
by the occurrence of BICs. N\"{o}ckel investigated theoretically the ballistic
transport across a quantum dot in a weak magnetic field.\cite{Nockel92}
Resonances in the transmission were found to grow narrower with decreasing the
magnetic field, and eventually they become BICs as the magnetic field vanishes. 
Fabry-P\'{e}rot interference of quasibound states of two open quantum dots
connected by a long wire cause the occurrence of BICs.\cite{Ordonez06} The
resulting state is nonlocal, in the sense that the electron is trapped in both
quantum dots at the same time. Interestingly, controlling the size of one of the
quantum dots makes the electron flows or get trapped inside the dots.  BICs in
parallel double quantum dot systems were found to be robust even if
electron-electron interaction is taken into account.\cite{Zitko12} Recently,
Gonz\'{a}lez~\emph{et al.\/} have demonstrated that not only quantum dots based
on semiconductor materials but also on graphene can support
BICs.\cite{Gonzalez10} All these features stimulate the interest of BICs to
develop new applications in nanoelectronics. 

In this work we extend the notion of BIC to the domain of time-dependent
potentials. Our aim is twofold. First, we explore the possibility of the
occurrence of BICs when the quantum system is driven by a time-harmonic
potential. Second, we introduce a physical realizable system which opens a novel
possibility to reveal the existence of these exotic states in transport
experiments. As a major result, we show that the transmission and 
correspondingly the conductance at low temperature signals the occurrence of
BICs as Fano resonances. When the system is driven by an AC field the BICs
survive and the existence is revealed by dynamic Fano resonances in the
transmission probability. Remarkably, it turns out that their energy can be
tuned by changing the frequency of the field. Therefore, the conductance at low
temperature presents a minimum when the BIC crosses the Fermi level by varying
the driving frequency.

\section{Quantum ring under an AC side-gate voltage}

\begin{figure}[ht]
\centerline{\includegraphics[width=0.6\columnwidth]{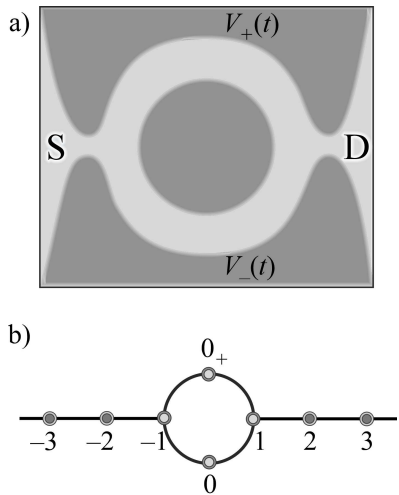}}
\caption{a)~Schematic diagram of the quantum ring with a side-gate voltage
$V_{\pm}(t)$ connected to source~(S) and drain~(D). b)~Equivalent lattice model
with two time-dependent site energies $\varepsilon_{\pm}(t)$ at sites  labeled
$0_{\pm}$ and two other sites with index $\pm 1$ attached to semi-infinite
chains.}
\label{fig1}
\end{figure}

The system under consideration is a two dimensional gas of noninteracting
electrons in a quantum ring, shown schematically in Fig.~\ref{fig1}a). The ring
is connected to two leads (source and drain). A side-gate voltage $V_{\pm}(t)$
breaks the symmetry of the upper and lower arms of the ring and acts as an
additional parameter for controlling the electric current, as recently suggested
for graphene-based nanorings.\cite{Munarriz11,Munarriz12} We assume that the
side-gate voltage can be modulated harmonically in time with frequency $\omega$.

In order to study electron transport across the quantum ring, we mapped it onto
a much simpler yet nontrivial lattice model, depicted in Fig.~\ref{fig1}b). We
replace the actual quantum ring by four sites of a lattice within the
tight-binding  approximation. Two sites ($0_{\pm}$) have time-dependent energies
$\varepsilon_{\pm}(t)$ and the other two sites, labeled $\pm 1$, are connected
to semi-infinite chains. Time-dependent site energies are given by
$\varepsilon_{\pm}(t)=\pm 2 \Delta \cos \omega t$. To avoid the profusion of 
free parameters, we assume a uniform transfer integral and vanishing site
energies except at sites $0_{\pm}$, without loosing generality. The common 
value of the transfer integral will be set as the unit of energy and we take
$\hbar=1$ throughout the paper.

\subsection{Time-independent side-gate voltage}

To gain insight into the possible occurrence of BICs in the system, we consider
the time-independent case by setting $\omega =0$ for the moment. An incoming
plane wave $\psi_j^{\mathrm{in}}(t)=\exp[i(kj-Et)]$, with energy $E=-2\cos k$
within the bands of the leads, will be partially transmitted in the form
$\psi_j^{\mathrm{tr}}(t)=t_0\exp[i(kj-Et)]$. The lattice period is set as the
length unit. It is a matter of simple algebra to obtain the transmission
amplitude in this case
\begin{equation}
t_0=\frac{4\sin k}{4\sin k+i\left(E+4\Delta^2/E\right)}\ .
\label{eq:1}
\end{equation}
The transmission probability $T(E)=|t_0|^2$ presents a dip around the band
center and vanishes at $E=0$. 

In the weak coupling limit, namely $\Delta \to 0$, the transmission probability
shows a Fano profile~\cite{Fano61} close to the band center,
$T(E)=E^2/\left(E^2+\Delta^4\right)$. The width of the dip scales as $\sim
\Delta^2$. Poles of the transmission amplitude have a simple physical
interpretation as the natural eigenstates of the scattering 
potential.\cite{Bagwell92} The poles occur at a  complex energy whose real part
gives the energy of the state and the imaginary part is related with its decay
rate. There are four poles in the case under study. Two poles correspond to
defect modes in the gap and, consequently, cannot be identified with BICs.
However, the other two poles reside at the band center with an imaginary part
equal to $\pm i \Delta^2$ when $\Delta \ll 1$. In the weak coupling limit
$\Delta\to 0$ the poles correspond to truly bound states. Therefore, the Fano
resonance of the  transmission amplitude~(\ref{eq:1}) signals the occurrence of
BICs at the band center.

To get a better understanding of the nature of the BICs at the band center we
also calculate the local density of states (LDOS) at sites $0_{\pm}$,
$\rho_0(E)$. Close to the band center, the LDOS is proportional to
$|\psi^{+}|^2+|\psi^{-}|^2$, where $\psi^{\pm}$ is the wave function amplitude
at those sites. The LDOS is given approximately as
\begin{equation}
\rho_{0}(E) \sim \frac{E^2}{E^2+\Delta^4}+\frac{4\Delta^2}{E^2+\Delta^4}\ .
\label{eq:2}
\end{equation}
The first term is nothing but the transmission probability, vanishing at the
band center. However, the second term approaches $4\pi\delta(E)$ in the limit
$\Delta\to 0$, indicating the existence of a truly bound state with energy $E=0$
located at sites $0_{\pm}$.

\subsection{Time-dependent side-gate voltage}

We now turn to our main goal, the occurrence of BICs when the side-gate voltage
depends harmonically on time. The time-dependent Schr\"{o}dinger equation for
the amplitudes $\psi_{j}(t)$ reads
\begin{equation}
i\dot{\psi}_j=\varepsilon_{\pm}(t)\delta_{j,0_{\pm}}\psi_{j}
-\sum_{i(j)}\psi_{i(j)} \ ,
\label{eq:3}
\end{equation} 
where the index $i(j)$ runs over the nearest-neighbor sites of $j$ and the dot
indicates the derivative with respect to time. Using the Floquet formalism, the
solution can be expressed in the form
\begin{subequations}
\begin{equation}
\psi_{j}(t)=\sum_{n=-\infty}^{\infty}A_{n,j}e^{-iE_{n}t}\ ,
\label{eq:4a}
\end{equation}
where $E_{n}=E+n\omega$ and $n$ is the sideband channel index. Since we are
interested in electron transmission  across the ring, we take the following
ansatz for the coefficients $A_{n,j}$ in the expansion~(\ref{eq:4a})
\begin{equation}
A_{n,j}=
\left\{
\begin{array}{ll}
\delta_{n0}e^{ik_{n}j}+r_{n}e^{-ik_{n}j}\ , & j\leq -1\ , \\
f_{n}^{\pm} \ , & j=0_{\pm}\ , \\
t_{n}e^{ik_{n}j}\ ,  & j\geq 1\ .
\end{array}
\right.
\label{eq:4b}
\end{equation}
\label{eq:4}
\end{subequations}
Inserting this ansatz in~(\ref{eq:3}) leads to the dispersion relation
$E_n=-2\cos k_{n}$ where $k_{n}$ is real if $E_{n}$ lies within the band, i.e.
$|E+n\omega|\leq 2$. In addition, we obtain
$t_{n}=f_{n}^{+}+f_{n}^{-}=r_{n}+\delta_{n0}$, ensuring current conservation.
Finally, one also gets
\begin{subequations}
\begin{equation}
\alpha_{n}t_{n} 
-\frac{\Delta^2}{E_{n-1}}\,t_{n-2}
-\frac{\Delta^2}{E_{n+1}}\,t_{n+2}=4i\delta_{n0}\sin k_{0}\ , 
\label{eq:5a}
\end{equation}
where for brevity we define
\begin{equation}
\alpha_{n} = 4i\sin k_{n}-E_{n}-\Delta^{2}
\left(\frac{1}{E_{n+1}}+\frac{1}{E_{n-1}}\right)\ .
\label{eq:5b}
\end{equation}
\label{eq:5}
\end{subequations}

The continued fraction approach developed in Ref.~\onlinecite{Martinez01} allows us
to obtain numerically the contribution of all channels to the transmission. But
if the coupling of the ring to the AC side-gate voltage is weak ($\Delta \ll
1$), only the lowest order sidebands are significant. Then we keep five channels
and assume that $t_{n}$ vanishes if $|n|\ge 3$. Equation~(\ref{eq:5}) implies
that $t_{\pm 1}=0$ in this approximation and 
\begin{subequations}
\begin{equation}
t_{\pm 2}=\frac{\Delta^2}{\alpha_{\pm 2}E_{\pm 1}}\,t_{0}\ . 
\label{eq:6a}
\end{equation}
The transmission amplitude in the elastic channel is
\begin{equation}
t_{0}=4i\sin k_{0}
\left[\alpha_{0}-\Delta^4\left(\frac{1}{\alpha_{2}E_{1}^2}+
\frac{1}{\alpha_{-2}E_{-1}^2}\right)\right]^{-1}\ .
\label{eq:6b}
\end{equation}
\label{eq:6}
\end{subequations}

Once the transmission amplitudes have been calculated, we can obtain the
transmission probability from the general expression
\begin{equation}
T_{\omega}(E)=\sum_{n} \frac{\sin k_{n}}{\sin k_{0}}\,|t_{n}|^2\ ,
\label{eq:7}
\end{equation}
where the sum runs over the propagating channels, namely those channels for
which $E_{n}=E+n\omega$ lies within the band of the leads.  Assuming that the
sidebands are active for conduction, the transmission probability reads
\begin{align}
T_{\omega}(E)&=|t_{0}|^2
\left(
1+\frac{\Delta^4}{|\alpha_{2}|^2E_{1}^2}\,\frac{\sin k_{2}}{\sin k_{0}}
\right. \nonumber\\
&\left.
+\frac{\Delta^4}{|\alpha_{-2}|^2E_{-1}^2}\,\frac{\sin k_{-2}}{\sin k_{0}}
\right) \ .
\label{eq:8}
\end{align}

\section{Results}

\begin{figure}[ht]
\centerline{\includegraphics[width=0.7\columnwidth,clip=]{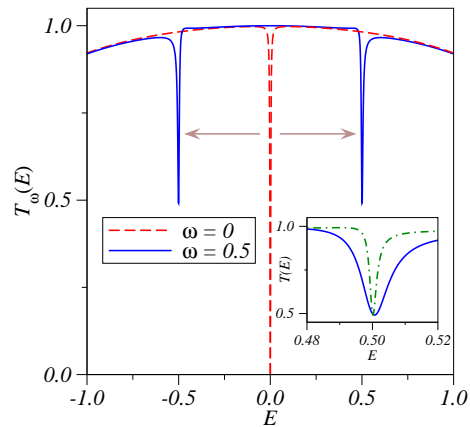}}
\caption{Transmission probability as a function of energy at $\omega=0$ (red
dashed line) and $\omega=0.5$ (blue solid line) for $\Delta=0.1$. The inset
shows an enlarged view of the transmission probability when $E \simeq \omega$ at
$\Delta=0.01$ (green dashed line) and $\Delta=0.1$ (blue solid line) for
$\omega=0.5$.} 
\label{fig2}
\end{figure}

The five channels approximation discussed above provides a closed analytical
expression for the transmission probability~(\ref{eq:8}). Admittedly the
resulting expression is still involved and must be evaluated numerically for the
chosen parameters. We will present some simpler expressions latter, valid in the
weak coupling limit, but for the moment we are interested in the closed 
expression~(\ref{eq:8}). We assume that coupling $\Delta$  is smaller or in the
order of the frequency $\omega$ hereafter. Figure~\ref{fig2} shows the results
when the frequency is $\omega=0.5$ and $\Delta=0.1$ (recall that the amplitude
is actually twice this value). The pronounced dip observed at the band center
in the static case ($\omega=0$), commented above  and shown in the figure, is
absent if $\omega\neq 0$. Similar results can be obtained for other set of
$\Delta$ and $\omega$ parameters. In fact, transmission at $E=0$ is unity and
the quantum ring becomes transparent at this energy. But the most salient
feature of the transmission when the side-gate voltage oscillates is the
occurrence of two symmetric and narrow dips, at energies close to $\pm \omega$.
Remarkably, the transmission never vanishes in the range of energy plotted in
Fig.~\ref{fig2} and at the dips only drops at about $0.5$. Actually transmission
probability vanishes but only at the band edges, as occurs in the static case
too. The inset shows an enlarged view of one of the dips for two different
values of the coupling $\Delta$. It is quite apparent that the minimum
transmission is slightly smaller than $0.5$ and it is reached at an energy close
but not exactly equal to $\omega$. In fact, transmission at $E=\omega$ is
exactly equal to $0.5$ for any value of $\Delta$. 

It is important to mention that we also solved numerically the general
equation~(\ref{eq:5}) to obtain the transmission probability from~(\ref{eq:7}),
increasing the number of the sideband channels. The results were the same to
those obtained within the five channels approximation, when $\Delta$ is not too
large. Thus, we can confidently use this approximation in our analysis.

\begin{figure}[ht]
\centerline{\includegraphics[width=0.7\columnwidth,clip=]{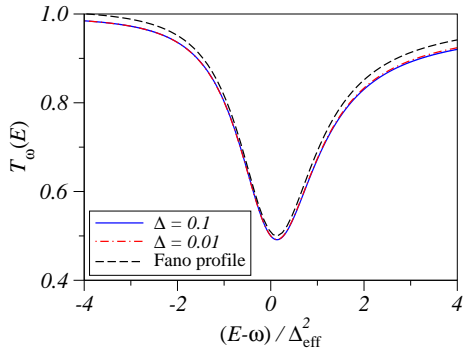}}
\caption{Transmission probability as a function of the parameter $x$  defined
in~\protect{(\ref{eq:9b})}, for two values of $\Delta$ and $\omega=0.5$. The
Fano profile is a good approximation to both curves.} 
\label{fig3}
\end{figure}

The transmission probability is an even function of energy. Therefore, for
concreteness we now focus in the energy region close $+\omega$, when $\Delta$ is
small. If $\omega$ is not large, we can take $\sin k_{\pm 2}\simeq \sin
k_{0}\simeq 1$ in~(\ref{eq:8}). In addition, the term containing $\alpha_{2}$ in
Eqs.~(\ref{eq:6b}) and~(\ref{eq:8}) is negligible under these assumptions. After
lengthly but straightforward algebra, the transmission probability reduces to
\begin{subequations}
\begin{equation}
T_{\omega}(E) \simeq \frac{8(E-\omega)^2-\omega(E-\omega)\Delta^2+\Delta^4}%
{8(E-\omega)^2+2\Delta^4}\ .
\label{eq:9a}
\end{equation}
Since $\omega$ and $E-\omega$ are not large, this expression can be further
approximated  by the following Fano profile
\begin{equation}
T_{\omega}(E) \simeq \frac{1}{2} + \frac{1}{2}\,\frac{(x-\omega/4)^2}{1+x^2}\ ,
\quad
x=\frac{E-\omega}{\Delta_{\mathrm{eff}}^{2}}\ .
\label{eq:9b}
\end{equation}
\label{eq:9}
\end{subequations}
To facilitate the comparison with the static case, we have introduced an
effective coupling $\Delta_{\mathrm{eff}}$ in such a way that
$\Delta_{\mathrm{eff}}^{2}=\overline{\varepsilon_{\pm}^2(t)}=\Delta^2/2$, where
the bar indicates the average over one time period. Notice that the Fano factor
$q=-\omega/4$ becomes independent of the coupling $\Delta$.  Figure~\ref{fig3}
compares the exact results obtained from~(\ref{eq:8}), for two different values
of the coupling $\Delta$ and $\omega=0.5$, and the Fano profile~(\ref{eq:9b}). 
In spite of the simplicity of the Fano profile and the assumptions we made, the
agreement is remarkable in both cases.

In contrast to the static case, transmission at the dips remains finite when the
side-gate voltage is harmonically modulated in time (see Fig.~\ref{fig2}).
Therefore, it is not clear at this stage whether the dips are due to the
occurrence of BICs in the system. To answer this question we consider again the
LDOS at sites $0_{\pm}$. After time-averaging over one time period, one gets
\begin{align*}
\overline{\rho_{0\omega}}(E) &\sim \sum_{n} 
\left(|f_{n}^{+}|^2+|f_{n}^{-}|^2\right) \\ 
&=\frac{1}{2}\sum_{n}\left(|t_{n}|^2
+\frac{\Delta^2}{E_{n}^{2}}\,|t_{n+1}+t_{n-1}|^2\right)\ .
\end{align*}
Using the five channels approximation, the LDOS close to right dip 
($E\ \sim \omega$) becomes 
\begin{equation}
\overline{\rho_{0\omega}}(E) \sim \frac{1}{2}\,T_{\omega}(E) 
+ \frac{\Delta_{\mathrm{eff}}^2}{(E-\omega)^2+\Delta_{\mathrm{eff}}^4}\ ,
\label{eq:10}
\end{equation}
and a similar expression is obtained for the left dip ($E\ \sim -\omega$), 
replacing $\omega$ by $-\omega$ in~(\ref{eq:10}). Therefore, in the weak
coupling limit $\Delta\to 0$, the LDOS reduces to  $\overline{\rho_{0\omega}}(E)
\sim \delta(E+\omega)+ \delta(E-\omega)$. In analogy with the static case, we
claim that the two singular peaks in the LDOS at energies $\pm \omega$ are due
to a new type of BICs arising by the interaction with the AC field.

\section{Conclusions}

In summary, we have introduced and studied a novel type of BICs in systems whose
energy levels are modulated harmonically in time. To be specific, we have
considered a quantum ring subjected to a side-gate voltage oscillating in time
with frequency $\omega$ and studied its transport properties in a fully coherent
regime. We come to the important conclusion that the BICs supported by the
quantum ring in the static case survive under harmonic modulation of the
side-gate voltage. The two BICs driven by the AC field have energies $\pm
\omega$ and they reveal themselves in the transmission, and consequently in the
low-temperature conductance, as dynamic Fano resonances. The position of the
BICs inside the spectral band can be continuously tuned by varying the driving
frequency and eventually they could be expelled out of the continuum when 
$\omega$ is larger than $2$ in units of the transfer integral, i.e. when they
approaches the band edge.  Similar control of the energy of static BICs have
been recently demonstrated by adding weak nonlinearity to semi-infinite
systems~\cite{Molina12} or by varying the interaction between
particles.\cite{Zhang12} However, besides the different origin of the BICs, our
proposal seems to be more advantageous for the experimental validation of these
exotic states. In this regard, it is still an open question to what extend
electron-electron  interactions would mask the effect in a real experiment.
\v{Z}tiko \emph{et al.} have shown that the so-called dark states in  parallel
double quantum dot systems are robust against interactions  within a Hubbard
model, at least in the Kondo regime.\cite{Zitko12} These states correspond to
the BICs of our present work in the static case, which makes us confident
to expect also the BICs driven by AC fields to be robust in an interacting
system.

\begin{acknowledgments}

Work at Madrid was supported by MICINN (project MAT2010-17180). C.\ G.-S.
acknowledges financial support from Comunidad de Madrid and European Social
Fundation. P.\ A.\ O. acknowledges financial support from FONDECYT under grant
1100560.

\end{acknowledgments}

\end{document}